\documentclass[sn-basic]{sn-jnl}

\usepackage{array}
\usepackage{placeins}
\usepackage{multicol}
\usepackage{graphicx}
\DeclareMathOperator{\argmax}{argmax}

\newcount\Comments  
\usepackage{color}
\definecolor{darkgreen}{rgb}{0,0.5,0}
\definecolor{purple}{rgb}{1,0,1}
\newcommand{\kibitz}[2]{\ifnum\Comments=1\textcolor{#1}{#2}\fi}

\usepackage{ftnright}
\usepackage[bottom]{footmisc}

\jyear{2021}%

\begin{document}

\title[Deep Reinforcement Trading with Predictable Returns]{Deep Reinforcement Trading with Predictable Returns}


\author*[1]{\fnm{Alessio} \sur{Brini}}\email{alessio.brini@duke.edu}

\author[2]{\fnm{Daniele} \sur{Tantari}}\email{daniele.tantari@unibo.it}

\affil*[1]{\orgname{Duke University}, \orgaddress{\city{Durham, NC}, \country{USA}}}

\affil[2]{\orgname{University of Bologna}, \orgaddress{ \city{Bologna}, \country{Italy}}}

\abstract{Classical portfolio optimization often requires forecasting asset returns and their corresponding variances in spite of the low signal-to-noise ratio provided in the financial markets. Modern deep reinforcement learning (DRL) offers a framework for optimizing sequential trader decisions  but lacks theoretical guarantees of convergence. On the other hand, the performances  on real financial trading problems are strongly affected by the goodness of the signal used to predict returns. To disentangle the effects coming from  return unpredictability  from those coming from algorithm un-trainability,  we investigate the performance of model-free DRL traders in a market environment  with different known mean-reverting factors driving the dynamics. When the framework admits an exact dynamic programming solution, we can assess the limits and capabilities of different value-based algorithms to retrieve meaningful trading signals in a data-driven manner.  We consider DRL agents that leverage classical strategies to increase their performances and we show that this approach guarantees flexibility, outperforming the benchmark strategy when the price dynamics is misspecified and some original assumptions on the market environment are violated with the presence of extreme events and volatility clustering.}




\maketitle

\section{Introduction}
The important milestone represented by the modern portfolio theory of \cite{M52} has set the basis for the beginning of financial portfolio optimization as an active field of research. Its original formulation suffers several drawbacks \citep{KTF2014} and has been extended from a single-period to a multi-period framework  to capture intertemporal effects and to allow  dynamical portfolio rebalancing  \citep{Grinold2006,Engle2007,Tut2011,kolm2012algorithmic,kolm2014multiperiod}. However, the addition of the time dimension makes even more complicated the estimation of an optimal strategy, which requires forecasting financial quantities such as risks and returns for several periods in the future. Single-period models are often still adopted because their dynamic counterpart is not practical and the forecasting step may lead to systematic errors due to the uncertainty about the chosen model or the inherent presence of a low signal-to-noise ratio in the financial data. Even when a multi-period model is effective in capturing the market impact or alpha decay, classical optimal control techniques lay over a set of restricting assumptions that cannot properly represent the real financial world.

In this work, we use reinforcement learning (RL) \citep{BarSut2018,Szepesvari2010} as a convenient framework to model sequential decision problems of a financial nature without the need to directly model the underlying asset dynamics. RL finds its roots in the optimal control theory along with the dynamic programming literature \citep{Bert05} and has gained a huge revival after the last decade's improvement of deep learning (DL) as a field of research. This gave rise to the so-called deep reinforcement learning (DRL) that has already obtained relevant results in application domains such as gaming \citep{silver2016mastering,mnih2015human} and robotics \citep{levine2016end}. For a comprehensive overview of DRL methods and their fields of application, see \cite{arulkumaran2017brief}.

The RL approach is not new to the financial domain and there are examples of practical applications for trading and portfolio management \citep{zhang2020deep,jiang2017deep}. However, recent DRL algorithms are very often homemade recipes without theoretical control. For this reason, the study of their performances in real financial trading problems is always an intricate combination of different effects, some of them related to the goodness of the dataset and the signals used to predict returns, some others related to the specific algorithm and its trainability issues.

To the best of our knowledge, there is a lack of research works that investigate  DRL performances in  financial trading problems besides the issues coming from market efficiency: the search for a good signal to predict returns or the possible lack of any signals in the dataset.
For this reason, we consider a controlled environment in which a signal is known to exist and study the capability
of DRL agents to discover profitable opportunities in the market, all while analyzing the role of the dataset structure against the algorithm architecture.

 Similarly to \cite{KR2019,chaouki2020deep}, we simulate financial asset returns which contain  predictable factors and we let the agent trade in an environment whose associated optimization problem admits an exact solution \citep{GarPed13}. The optimal benchmark strategy allows us to evaluate the strengths and flaws of a DRL approach, both in terms of accuracy and efficiency.  

As the main novelty of our work, we exploit a data-driven setting of DRL in which the agents not only compete against classical strategies but can also leverage their experience to optimize the state-action space and increase the learning speed. We test different DRL approaches on a variety of financial data with different properties to investigate their flexibility when the simulated dynamics is misspecified with respect to the assumptions of the benchmark model.  
We  show that model free DRL algorithms can reach the performance of the benchmark strategy, when it is optimal, and also outperform it in the case of  model misspecifications like the presence of extreme events and volatility clustering.

This opens the possibility of using DRL, not only directly on real data, but also as a tool for searching optimal strategies of rich and realistic (straying from the efficient market hypothesis) generative models of financial markets, where time and data scarcity is not an issue.

We also show that classical strategies can help DRL agents by giving them information about the typical scale of a good strategy to start and adjust.


\section{Financial Market Environment}\label{mkt_env}
The agent operates in a financial market where at each time $t\in\mathbb{Z}$ it can trade N securities whose excess returns $y_{t+1} = p_{t+1} - (1 + r_{f}) p_{t}$ are given by 
\begin{equation}\label{Eq:returns}
    y_{t+1} = Bf_{t} + u_{t+1},
\end{equation}
where $f_{t}$ is a $K\times 1$ vector of return-predicting factors, $B$ is a matrix of factor loadings and $u_{t+1}$ is a noise term with $\mathbb{E}[u_{t+1}]=0$ and $\operatorname{Var}[u_{t+1}] = \Sigma$. 

The factors can be either value factors, which describe the profitability of the asset relative to some fundamental measure, or momentum factors, which rely on past price movements to predict the future \citep{tsmom_moskovitz}. We assume they evolve according to a discretization of a mean-reverting process \citep{OU}
\begin{equation}\label{Eq:factors}
    \Delta f_{t+1} = - \phi f_{t} + \epsilon_{t+1},
\end{equation}
where $\phi$ is a $K\times K$ matrix of mean-reversion coefficients and $\epsilon_{t+1}$ represents a stochastic shock component  with $\mathbb{E}[\epsilon_{t+1}]=0$ and $Var[\epsilon_{t+1}] = \Omega$. 

Trading in this environment produces transaction costs which we assume to be a quadratic function of the traded amount $\Delta h_{t} = h_{t} - h_{t-1}$, i.e.
\begin{equation}\label{Eq:tcosts}
    C(\Delta h_{t}) = \frac{1}{2}\Delta h_{t}^{T} \Lambda \Delta h_{t},
\end{equation}
where $\Lambda$ is a symmetric positive definite matrix ensuring transaction costs convexity as generally required by empirical literature \citep{lillo2003master,garleanu2008demand} and is consistent with the assumption of a linear price impact. In the following, we also assume that $\Lambda = \lambda \Sigma$, i.e. that trading costs are actually the compensation for the dealer's risk that takes the other part of the transaction. In this context, $\lambda$ can be interpreted as the dealer's risk aversion and controls the degree of liquidity of the asset.

The agent's goal is to find a dynamic portfolio strategy $(h_{0},h_{1},\ldots)$ by maximizing the present value of all future returns, penalized for risk and net of transaction costs, i.e.
\begin{equation}\label{objective}
    \max_{(h_{0},h_{1},\ldots)} \mathbb{E}_{0} \Big[ \sum\limits_{t}\rho^{t+1} (h_{t}^{T} y_{t+1} - \frac{\gamma}{2} h_{t}^{T}\Sigma h_{t}) - 
    +  \frac{\rho^{t}}{2} \Delta h_{t}^{T} \Lambda \Delta h_{t} \Big],
\end{equation}%
where $\rho \in (0,1)$ is a discount rate and $\gamma$ is the risk aversion coefficient.

When the noise terms $u_t$ and $\epsilon_t$ are assumed to be distributed as a Gaussian, the model coincides with  \cite{GarPed13} that has a closed-form solution as
\begin{equation}\label{optportfolio}
    h_{t} = \left( 1 - \frac{a}{\lambda}\right) h_{t-1} + \frac{a}{\lambda}h_{t}^{aim},
\end{equation}
i.e. the optimal strategy is  a convex combination of holding the previous portfolio and trading towards the objective portfolio $h_t^{aim}$ with a trading rate $a/\lambda$. Recall that $\lambda$ is a scalar parameter that controls the magnitude of the transaction costs matrix $\Lambda = \lambda \Sigma$.  The trading rate  $\frac{a}{\lambda} < 1$, where
\begin{equation}
a = \frac{-(\gamma(1 - r_{f}) + \lambda r_{f}) + \sqrt{(\gamma(1 - r_{f}) + \lambda r_{f})^{2} + 4\gamma \lambda (1 - r_{f})^{2}}}{2(1 - r_{f})}
\end{equation}
is a decreasing function of the transaction costs by the effect of $\lambda$ and increasing in the risk aversion $\gamma$. The objective portfolio $h_t^{aim}$ is defined by 
 \begin{equation}
    h_{t}^{aim} = (\gamma \Sigma)^{-1} B(I + \frac{a}{\gamma} \Phi)^{-1} f_{t},
\end{equation}
where $B$ and $\Phi$ are defined in Eqs. (\ref{Eq:returns}),(\ref{Eq:factors}). It is a generalization of the well-known Markovitz portfolio  \citep{M52}
\begin{equation}
    h_{t}^{M} = \left( \gamma \Sigma \right)^{-1} Bf_{t},
\end{equation}
which is optimal only in the static case and in absence of transaction costs. Instead, the aim portfolio in eq. (\ref{optportfolio}) represents a dynamic strategy and can be shown to be a weighted average of all future Markovitz portfolios.

If the mean reversion coefficient matrix $\Phi$ is diagonal, the aim portfolio become  
\begin{equation}\label{Eq:aim_port}
    h_{t}^{aim} = (\gamma \Sigma)^{-1} B\left(\frac{f_{t}^{1}}{1 + \Phi^{1} \frac{a}{\gamma}}, \ldots, \frac{f_{t}^{K}}{1 + \Phi^{K} \frac{a}{\gamma}}\right)^{T},
\end{equation}
where the K factors are scaled down by their speed of mean-reversion $\Phi$.  A factor $i$ with a slower speed of mean-reversion is scaled less than a faster factor $j$ and the relative weight of $f^{i}$ with respect to the weight of $f^{j}$,  
$
    \frac{1 + \Phi^{j} \frac{a}{\gamma}}{1 + \Phi^{i} \frac{a}{\gamma}}
$
increases with the transaction cost $\lambda$. In fact, the cost friction leads the investor to slow down the rate of portfolio rebalancing and faster factors require to close out the position in a shorter time frame.   

In the following, we will use the optimal strategy (\ref{optportfolio}) of the Gaussian model as a benchmark for the DRL performance in solving the problem (\ref{objective}) but we  also consider other possible model specifications for which an explicit optimal solution is not available. In particular, we introduce fat-tailed distributed shocks and heteroskedastic volatility as interesting model misspecifications that reflect general properties of empirical asset returns. \citep{cont2001empirical}.

A riskier environment with many extreme events is constructed by assuming the asset noise to depart from a Gaussian distribution. In particular, we consider $u_t$ and $\epsilon_t$ distributed as a Student's T distribution with $\nu$ degrees of freedom. On the other hand, heteroskedasticity  is introduced  according to a generalized autoregressive conditional heteroskedastic (GARCH) process \citep{bollerslev1987conditionally}  for the variance of asset returns to model volatility clustering. In the case of a single asset it means that $u_{t}=\sigma_{t} z_{t}$ where
\begin{equation}\label{Eq:garch}
\sigma_{t}^{2} = \omega+\sum\limits_{j=1}^{p} \alpha_{j}\mid u_{j-i} \mid ^{2}+\sum\limits_{k=1}^{q} \beta_{k} \sigma_{t-k}^{2} 
\end{equation}
and $z_t$ is a noise term that can be either a standard Gaussian or a Student's T with $\nu$ degrees of freedom.


\section{Deep Reinforcement Learning Methods}\label{RL}
The aim of RL is to solve a decision-making problem in which the timing of costs and benefits is relevant. Financial portfolio optimization comprises a set of problems where current actions can influence the future, even at a very distant point in time. RL approaches the resolution of this problem by trial and error, learning by obtaining a feedback after each sequential decision. 

An RL problem can be formulated in the context of a Markov Decision Process (MDP), which is defined by a set of possible states  $S_t\in\mathcal{S}$, a set of possible actions $A_t\in\mathcal{A}$ and a transition probability  $\mathcal{P}_{s s'}^{a}=P[S_{t+1} = s' \mid S_{t} = s, A_{t} = a]$. Therefore, it is the (stochastic) control problem of finding
\begin{equation}\label{Eq:rlproblem}
\max_{\{\pi\}} \mathbb{E}\left[ \sum_{t=0}^\infty \rho^t R_{t+1}(S_t,A_t,S_{t+1} ) \right] 
\end{equation}
where $\pi$ defines the agent's strategy that associates a probability $\pi(a \mid s)$ to the action $A_{t}=a$ given the state  $S_{t}=s$. An RL agent aims at maximizing the expected sum of (discounted) rewards by finding the best action given the current state.
We consider the model-free context in which the agent has no knowledge of the internal dynamics of the environment, i.e. the transition probability is not known, and the only source of information is the sequence of states, actions, and rewards.

Value-based methods are defined by  introducing the action-value function
\begin{equation}\label{Eq:Qfunc}
Q_{\pi}(s, a) \equiv \mathbb{E}\left[\sum_{k=0}^\infty  \rho^k R_{t+1+k} \mid S_{t}=s, A_{t}=a, \pi\right], 
\end{equation}
which reflects the long-term reward associated with the action $a$ taken in the state $s$ if the strategy $\pi$ is followed hereafter. The estimation of (\ref{Eq:Qfunc}) allows to derive a deterministic optimal policy
as the highest valued action in each state.
Depending on how the  agent estimates the action-value function (\ref{Eq:Qfunc}), different classes of value-based algorithms can be introduced.
Conversely, direct policy search approaches are alternative methods that try to explore directly  the policy space (or some subset of it), being  the problem a particular case of stochastic optimization.

\subsection{Tabular Reinforcement Learning}
Tabular RL methods are practical when the possible  states and  actions are few enough to be represented in a table, which has an entry for every $(s,a)$ pair. In this case, the agent can explore many possible state-action pairs within a reasonable amount of computational time and obtain a good approximation of the value function.  

Q-learning \citep{Watkins92q-learning} is a tabular method in which at each timestep the agent tries an action $A_t$, receives a reward $R_{t+1}$ and updates the current estimate of the action-value function $Q(S_{t},A_{t})$ as
\begin{equation}\label{Eq:Qlearning}
    Q(S_{t},A_{t}) \gets Q(S_{t},A_{t}) + \alpha (T^Q_t-Q(S_{t},A_{t})),
\end{equation}
where $\alpha$ is a learning rate and the target \begin{equation}
    T^Q_{t} = R_{t+1} + \rho \max_{a}Q(S_{t+1},a)
\end{equation}
is a decomposition of the value function in terms of the current reward  and the current estimate of the future value discounted by $\rho$. At the end of the learning process, the optimal policy is the greedy strategy  $A_{t}=\argmax_{a} Q(S_{t}, a)$ but Q-learning is trained off-policy because the agent chooses the action $A_t$ following an $\epsilon$-greedy policy that ensures adequate exploration of the state-action space.

When states and actions are continuous, as it is for a realistic financial environment, Q-learning barely obtains good estimates of the value function in a feasible computational time. Moreover, the discretization of the state space itself may cause a loss of relevant information depending on its granularity. In this context, the DRL framework becomes particularly necessary. 
\subsection{Approximate Reinforcement Learning}
DRL algorithms tackle previously intractable problems by approximating eq. (\ref{Eq:Qfunc}) through a neural network that allows a continuous state space representation. 

Deep Q-Network (DQN) \citep{mnih2015human} is an extension of Q-learning and allows learning a parametrized value function $Q^{*}(s, a)  \approx Q(s, a ; \theta)$. $Q(s, a ; \theta)$ is a multi-layer neural network that for a given input state $s$ returns a vector of action values. The standard update of eq. (\ref{Eq:Qlearning}) therefore becomes
\begin{equation}\label{DQLupdate}
    \theta_{t+1}=\theta_t +\alpha (T^{\text{DQN}}_t-Q(S_t, A_t ; \theta_t))\nabla_{\theta_t}Q(S_t, A_t ; \theta_t)
\end{equation}
which resembles a standard gradient descent toward the target 
\begin{equation}\label{Eq:DQNtgt}
T^{\text{DQN}}_t=R_{t+1} + \rho \max_{a}Q(S_{t+1},a;\theta_t).
\end{equation}
Even if tabular methods converge to the optimal function \citep{Watkins92q-learning}, they fail to generalize over previously unseen states. Instead, DRL has good generalization capabilities, but produces unstable behaviors during the training whenever function approximation is combined with an off-policy algorithm and learning by estimates \citep{BarSut2018}. 
The issue of training instability is partially solved by adding two ingredients: an experience replay buffer and a fixed target. An experience buffer is a finite set $\mathcal{D}=\{e_{1}, \ldots, e_{N}\}$ of fixed cardinality $N$, where at each time $t$ the agent's stream of experience $e_{t} = (S_{t}, A_{t}, R_{t+1}, S_{t+1})$ is stored replacing one of the old ones. The replay buffer is then used  to perform a batch update of the network parameters.
A fixed target is exactly as the online target except that its parameters $\theta^-$ are updated  ($\theta^-_t=\theta_t$) and then kept fixed for $\tau$ iterations. Combining the two ingredients the gradient step  of eq. (\ref{DQLupdate}) becomes
\begin{align}
   \mathbb{E}_{e}[(r+\rho \max _{a^{\prime}} Q\left(s^{\prime}, a^{\prime} ; \theta^-_t\right)
- Q\left(s, a ; \theta_t\right)) \nabla_{\theta_{t}} Q(s, a ; \theta_t)]
\end{align}
where $e=\left(s, a, r, s^{\prime}\right)$ is uniformly sampled from $\mathcal{D}$.

In what follows we adopt a variant of the algorithm called double DQN (DDQN) \citep{hasselt2015doubledqn} which prevents some overestimation issues of the value function. For convenience, we still refer to the chosen value-based algorithm as DQN, even if the implementation follows the variant of DDQN. In the appendix, we recap the technical details of the value-based algorithms used in the numerical experiments.

The optimization problem in eq. (\ref{Eq:rlproblem}) can be equivalently solved using a policy gradient algorithm like the Proximal Policy Optimization (PPO) \citep{SchulmanWDRK17}. A policy gradient algorithm directly parametrizes the optimal strategy within a given policy class $\pi_{\theta} = \pi(A_{t}\mid S_{t}; \theta)$, for example, a multilayer neural network with the set of parameters $\theta$. 
The optimization problem is approximately solved by computing the gradient of the performance measure  $J(\theta) = \sum_{t=0}^\infty \rho^t R_{t+1}(S_t,A_t,S_{t+1};\pi_{\theta})$ and then carrying out gradient ascent updates according to 
\begin{equation}\label{Eq:PGupdate}
    \theta_{t+1}=\theta_t +\alpha \nabla_{\theta} J(\theta_{t}), 
\end{equation}
where $\alpha$ is still a scalar learning rate. The policy gradient theorem \citep{sutton2000,marbach2001} provides an analytical expression for the gradient of $J(\theta)$ as
\begin{align}\label{Eq:pgtheorem}
    \nabla_{\theta} J(\theta)  &= \mathbb{E}_{\pi_{\theta}} \left[ \frac{\nabla_{\theta} \pi\left(A_{t} \mid S_{t};\theta \right)}{\pi\left(A_{t} \mid S_{t};\theta \right)}  Q_{\pi_\theta}(S_{t}, A_{t}) \right]  \\ \nonumber
    &= \mathbb{E}_{\pi_{\theta}} \left[  \nabla_{\theta} \log\pi\left(A_{t} \mid S_{t};\theta \right)  Q_{\pi_\theta}(S_{t}, A_{t}) \right],
\end{align}
where the expectation, w.r.t. $(S_t,A_t)$,  is taken along a trajectory (episode) that occurs adopting the policy $\pi_{\theta}$.
It can be proven that it is possible to modify the action value function $Q_\pi(s,a)$ in ($\ref{Eq:pgtheorem}$) by subtracting a baseline $V_\pi(s)$ that reduces the variance of the empirical average along the episode, while keeping the mean unchanged. A popular baseline choice is the state-value function
\begin{equation}\label{Eq:Vfunc}
    V_{\pi}(s) \equiv  \mathbb{E}\left[\sum_{k=0}^\infty  \rho^k R_{t+1+k} \mid S_{t}=s, \pi\right], 
\end{equation}
which reflects the long-term reward starting from the state $s$ if the strategy $\pi$ is adopted onwards. The gradient thus can be  rewritten as
\begin{equation}\label{Eq:pgadvantage}
    \nabla_{\theta} J(\theta)  = \mathbb{E}_{\pi_\theta} \left[  \nabla_{\theta} \log\pi\left(A_{t} \mid S_{t};\theta_{t} \right)  \mathbb{A}_{\pi\theta}(S_{t}, A_{t}) \right]
\end{equation}
where 
\begin{equation}\label{Eq:Afunc}
    \mathbb{A}_{\pi}(s, a) \equiv Q_{\pi}(s, a)-V_{\pi}(s),
\end{equation}
is called  advantage function and quantifies the  gain obtained by choosing a specific action in a  given  state with respect to its average value for the policy $\pi$.

Different policy gradient algorithms depend on how the advantage function is estimated. In PPO the advantage estimator  $\mathbb{A}\left(s,a;\psi\right)$ is parametrized  by another neural network with  parameters $\psi$. This approach is known as actor-critic: the actor is represented by the  policy estimator $\pi(a \mid s;\theta)$ that outputs the mean and the standard deviation of a Gaussian distribution which the agent uses to sample actions, the critic is the advantage function estimator $\mathbb{A}\left(s,a;\psi\right)$ whose output is a single scalar value. The two neural networks interact during the learning process: the critic drives the updates of the actor which successively collects new sample sequences  that will be used to update the critic and again evaluated by it for  new updates. The PPO algorithm can therefore be described by the extended objective function 
\begin{equation}\label{Eq:ppoobj}
    J^{\text{PPO}}(\theta,\psi)=  J(\theta)-c_{1} L^{\text{AF}}(\psi) +c_{2} H\left(\pi\left(a \mid s; \theta\right)\right).
\end{equation}
The second term is a loss  between the advantage function estimator $\mathbb{A}\left(s,a;\psi\right)$ and a target $\mathbb{A}^{targ}$, represented by the cumulative sum of discounted reward, needed to train the critic neural network.  The  last term represents an entropy bonus to guarantee an adequate level of exploration.  Details about the specific choice of the losses, the target, and the neural network parametrization, together with additional information  about the general algorithm implementation are given in the appendix. 

\section{Numerical Experiments}\label{exp}
In this section, we conduct synthetic experiments in the controlled financial environment outlined in Section \ref{mkt_env}. We present two different groups of experiments where the agents observe financial time series that come from different data-generating processes. The first group is related to the case where the return dynamics is driven by Gaussian mean reverting factors as in eq. (\ref{Eq:factors}) and the  optimal strategy is known to be eq. (\ref{optportfolio}). The second group includes a set of cases where the model that generates the dynamics allows for a bigger amount of extreme events and heteroskedastic volatility. In this case  eq. (\ref{optportfolio}) is still used  as a representative classical strategy of dynamic portfolio optimization.

In all experiments the agents trade a single asset, but the framework is general enough to allow for multi-asset trading. We test Q-learning and DQN in parallel in the same environment, while PPO training is slightly different. The first two algorithms are trained in-sample for a number of updates equal to the length $T_{in}$ of the simulated series, while the learned policy is evaluated out-of-sample at several intermediate training moments on different series of length $T_{out}$. The same logic operates for PPO, which however works in an episodic way: the algorithm is trained in-sample and evaluated out-of-sample respectively for $E_{in}$ and $E_{out}$ number of episodes of length equal to 2000 timesteps. Each agent operates in a model-free context so that no prior information about the data-generating process is provided. 

In order to bring the RL formalism to the portfolio optimization problem of eq. (\ref{objective}) we choose the actions as the amount of shares traded $A_t=\Delta h_t$, while the state is defined as the pair return-holding $S_t = (y_t,h_{t-1})$. We include the asset return in the state representation instead of the predicting factors because it is our interest to assess DRL as a purely data-driven approach. The choice of financial factors is known to be a non-trivial task and it can be highly discretionary. For every experiment, we also adapt the boundaries of the action space $\mathcal{A}$ according to the magnitude of the action performed by the benchmark. More specific details about this heuristic are provided in the appendix.

After taking an action and causing a change in portfolio position,  the agent observes the next price movement and the reward signal that is 
\begin{equation}\label{Eq:reward}
R_{t+1}(y_{t+1},h_{t-1},\Delta h_{t}) = h_{t}^{T} y_{t+1} - \frac{\gamma}{2} h_{t}^{T}\Sigma h_{t} - \frac{1}{2} \Delta h_{t}^{T} \Lambda \Delta h_{t}.
\end{equation}
Note that we decided to allow the benchmark agent to be perfectly informed so that it knows exactly the predicting factors of the price dynamics. On the contrary, RL agents can just gather information from the observed return which is affected by an additional source of noise. This choice allows the RL agent to be completely agnostic with respect to the price dynamics. This represents a clear disadvantage for the RL agent, but it is a step towards a more flexible approach when the dynamics is not known and the performance is strongly dependent on the selection of the factors. Alternatively one can assume that the RL agents are completely informed by replacing $y_t$ with $f_t$ in the definition of the state variables.
For the purpose of  comparison, we highlight that the value-based algorithms in this study perform discrete control, while the benchmark solution can adopt a continuous strategy according to eq. (\ref{optportfolio}). Although PPO can express both discrete and continuous policies, we test the continuous version to allow for a more expressive policy and compare the differences between the two settings.

For all the tested algorithms, we choose two performance metrics to evaluate the trading results of each RL agent from different perspectives: the net monetary gain on the one hand and the riskiness on the other. Therefore we compute the cumulative net $PnL$, which is expressed as the gross return of the portfolio deducted from the transaction costs, i.e.
\begin{equation}\label{eq:pnl_metrics}
    PnL_{t+1}^{\text{net}}(y_{t+1},h_{t},\Delta h_{t}) = h_{t} y_{t+1}  - \frac{1}{2} \Delta h_{t} \Lambda \Delta h_{t},
\end{equation}
and the annualized Sharpe ratio ($SR$)  \citep{sharpe1994sharpe} which is computed as
\begin{equation}\label{eq:sr_metrics}
    SR = \frac{\mathbb{E}[PnL^{\text{net}}]}{\sqrt{\operatorname{Var}[PnL^{\text{net}}]}} * \sqrt{252},
\end{equation}
defining the expected return of the portfolio per unit of risk on a yearly basis. It is a common metric to evaluate the trade-off between risk and return of financial strategies, especially in a mean-variance optimization framework.

The details about the parameters used to simulate the financial data and the hyperparameters setting for training the neural networks are provided in the appendix. The experiments are run in parallel on a 64-cores Linux server which has an Intel Xeon CPU E5-2683 v4 @ 2.10GHz. The training runtime for a single value-based method experiment of length $T_{in}=300000$ ranges from two to four hours when the neural network architecture is not deeper than two hidden layers. Approximately the same runtime is needed for the PPO algorithm when $E_{in}=300$. The source code written in Python is available on GitHub\footnote{https://github.com/Alessiobrini/Deep-Reinforcement-Trading-with-Predictable-Returns}. The following subsections discuss the results of the two groups of experiments.
\begin{figure}[tb]
  \centering\includegraphics{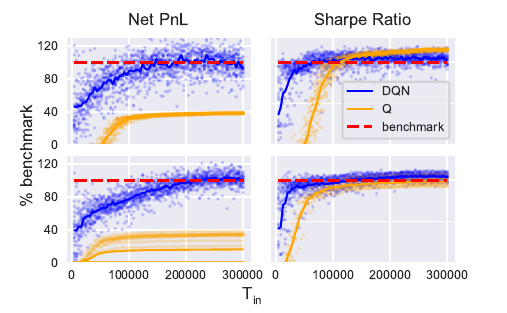}
  \caption{Results for DQN and Q-learning in the case of Gaussian dynamics driven by one (first row) and two (second row) mean-reverting factors. Cumulative net $PnL$ (first column) and $SR$ (second column) are displayed as the size of the training time series increases up to $T_{in}=300000$ on the x-axis. Every dot represents the average over $10$ out-of-sample tests of length $T_{out}=5000$ for a specific agent out of the 20 tested in total. The horizontal dashed line represents the optimal benchmark, while the solid lines represent the average performance of all the agents in relative percentage to the benchmark.}
  \label{Fig:gauss}
\end{figure}%
\subsection{Tracking the Benchmark}\label{SubSec:track_bench}
\begin{figure}[tb] 
  \centering\includegraphics{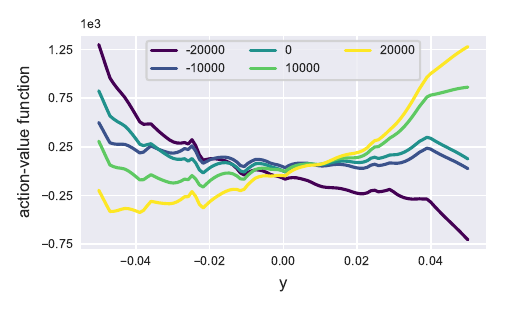}
  \caption{Learned action-value function for a DQN agent when the asset return varies and the holding is fixed at 0. Different colors represent different actions.}
  \label{Fig:valuefunc}
\end{figure}%

Figure \ref{Fig:gauss} presents the evolution of the out-of-sample performance of several Q-learning and DQN agents when the dynamics is driven by one (first row) and two (second row) mean-reverting factors. The first column displays  the cumulative net $PnL$ metric (eq. \ref{eq:pnl_metrics}), while the second column displays the $SR$ metric (eq. \ref{eq:sr_metrics}), as the size of the training time series increases up to Tin = 300000 on the x-axis. Every dot in Figure \ref{Fig:gauss} represents the average over $10$ out-of-sample tests of length $T_{out}=5000$ for a specific agent out of the 20 tested in total. The horizontal dashed line represents the optimal benchmark, while the solid lines represent the average performance of all the agents in relative percentage to the benchmark.

From Figure \ref{Fig:gauss} we observe that after about half of the training runtime, DQN reaches on average a close-to-optimal cumulative net $PnL$. The trained DQN agents are then able to retrieve the mean reverting signals in the data and control the number of transaction costs without knowing the data-generating process of the underlying dynamics. On the other hand, Q-learning agents hardly reach half of the cumulative net $PnL$ of the optimal benchmark in the same training time.

The performances of the tabular algorithm are strictly dependent on the granularity of the state discretization. Q-learning can reach the benchmark performance only when $T_{in} \to\infty$ and $\mathcal{S}$ are dense enough to closely represent the continuous trading environment. However, even if we keep the Q-table relatively  small in size, usually below 100000,  it can still be very sparse for this range of $T_{in}$. This is particularly evident in the case of two Gaussian factors, where many agents have a negligible cumulative net $PnL$ simply because they do not perform any buy or sell actions. Increasing the size of the Q-table for experiments of fixed length $T_{in}$ leads to even worse results. 


DQN avoids the inefficient tabular parametrization of the action-value function by using fewer parameters with respect to the number of entries in the Q-table. The use of a neural network as an action-value function approximator is crucial in this financial environment because the agent learns faster when the state space is entirely observable and the parameters can be updated by batches of experience. 

\begin{figure}[tb]
  \centering\includegraphics{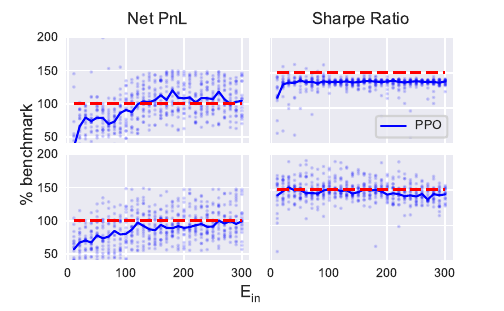}
  \caption{Results for PPO in the case of Gaussian dynamics driven by one (first row) and two (second row) mean-reverting factors. Cumulative net $PnL$ (first column) and $SR$ (second column) are displayed as the training episodes increase up to $E_{in}=300$ on the x-axis. Every dot represents the average over $10$ out-of-sample tests of length $T_{out}=2000$ for a specific agent out of the 20 tested in total. The horizontal dashed line represents the optimal benchmark, while the solid line represents the average performance of all the agents in relative percentage to the benchmark.}
  \label{Fig:gauss_ppo}
\end{figure}%
The second column of Figure \ref{Fig:gauss} showcases the evolution of the $SR$ of the agents and highlights that DQN obtains on average the same level of benchmark profit adjusted for risk since the beginning of the training. The performances of Q-learning, in this case, are strongly biased, since often the tabular agents choose not to trade and avoid increasing the risk of their portfolio position. The DQN agents first learn how to obtain low-risk portfolios, then they start making higher profits, as it is shown by the faster convergence of the $SR$ with respect to the cumulative net $PnL$.

\begin{figure}[tb] 
  \centering\includegraphics{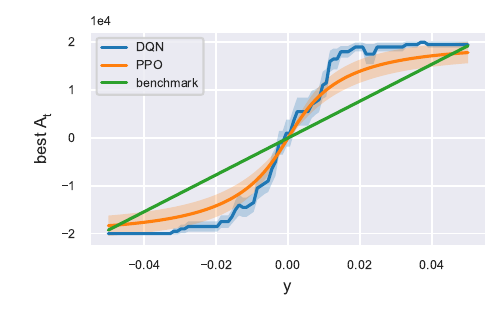}
  \caption{Greedy policy function for a DQN and a PPO agent together with the policy of the benchmark when the asset return varies and the holding is fixed at 0. Different colors represent different algorithms. The curves present a confidence interval because the average maximum action is obtained from the results of 20 trained agents for each algorithm.}
  \label{Fig:policies}
\end{figure}%
Figure \ref{Fig:valuefunc} provides insights about the learned behavior of the DRL agents showing the learned action-value function of the best performing DQN agent at the end of the period of training represented in Figure \ref{Fig:gauss}. The agent is trained over a return dynamics driven by one predicting factor, but the findings are valid also in the case of multiple factors. The estimated action-value function $Q((y,h),a;\theta)$ is displayed for all the actions in the discrete space $\mathcal{A}$ and implicitly represents the behavior of the agent when different levels of returns are experienced. If the agent acts greedily and chooses the highest Q-value for every level of $y$, positive actions appear prevalent when returns are positive, while the opposite holds for negative actions. 

Figure \ref{Fig:gauss_ppo} shows analogous experiment results for PPO trained with financial returns driven by mean-reverting Gaussian dynamics. The average out of sample performances of a representative agent is displayed as the number of training episodes increases. PPO retrieves the signal in the data and converges to the benchmark, but exhibits higher variance in the Net PnL measure compared to DQN. This is motivated by the different types of policies that the algorithm is describing. Working in a continuous action space allows the possibility to trade any fraction of the synthetic asset, but also complicates the exact convergence to the benchmark because the sampling space is large. In practice, there is no theoretical guarantee to find the proper way to sample actions from $\mathcal{A}$ in a finite time. 

Figure \ref{Fig:policies} represents the average greedy policy learned by the agents for DQN and PPO presented in Figures \ref{Fig:gauss} and \ref{Fig:gauss_ppo}. Both algorithms can discover the inherent arbitrage introduced in the market since the average policy follows the sign of the returns by buying low and selling high. The learned policies appear to be monotonic as one of the benchmarks. 
\subsection{Outperforming the Benchmark}
\begin{figure}[tb]
  \centering\includegraphics{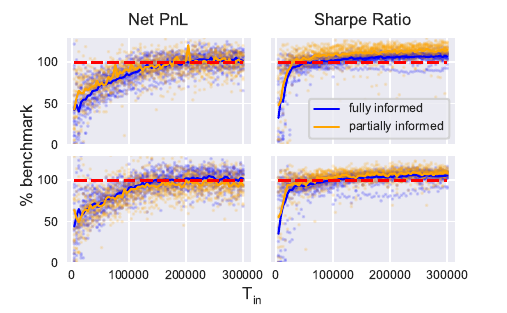}
  \caption{Results for Student's T dynamics in the case of one mean-reverting factor. DQN is tested over Student's T distributed returns with $\nu=6$ (first row) and $\nu=8$ (second row) respectively. The figure should be read with the same logic as Figure \ref{Fig:gauss} since the number of agents and the length of in-sample and out-of-sample experiments are equal. The solid lines represent different PPO performances with respect to different  benchmark strategies, which  in this context are not optimal anymore.}
  \label{Fig:stud}
\end{figure}%

\begin{figure}[tb]
  \centering\includegraphics{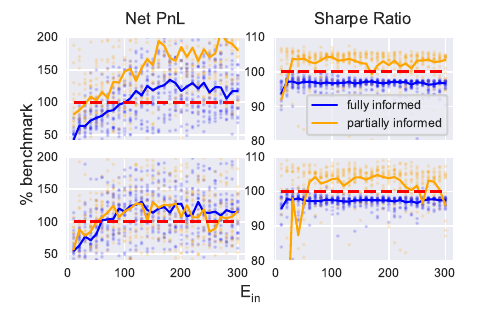}
  \caption{Results for Student's T dynamics in the case of one mean-reverting factor. PPO is tested over Student's T distributed returns with $\nu=6$ (first row) and $\nu=8$ (second row) respectively. The figure should be read with the same logic as Figure \ref{Fig:gauss_ppo} since the number of agents and the length of in-sample and out-of-sample experiments are equal. The solid lines represent different PPO performances with respect to different  benchmark strategies, which  in this context are not optimal anymore.}
  \label{Fig:stud_ppo}
\end{figure}%
In order to show the flexibility of the DRL approach, we study its performances with respect to the benchmark when the return dynamics depart from the original model specification.  In particular, we introduce two types of model misspecifications: the presence of extreme events and noise heteroskedasticity. In both cases the strategy in eq. (\ref{optportfolio}) is no more optimal, but since it performs well and is often used in practice, it can be considered as a benchmark representing a broader class of factor trading strategies. It is therefore natural to investigate whether DQN and PPO are able to outperform the benchmark other than just reaching it. 

We consider two different reference strategies that are respectively referred to as fully informed when the benchmark is provided with the simulated factors and partially informed when instead it needs to extract them from the observed returns. These different settings should not affect the DRL performance, except for the boundaries of the action space $\mathcal{A}$ that we decided to adapt to the benchmark for better comparison (see the appendix). 

The fully informed benchmark agent can directly use eq. (\ref{optportfolio}) to trade just by estimating the speeds of mean reversion and factor loadings from the observed return predicting factors. Instead, in the partially informed case, the benchmark agent does not know exactly which are the best predicting factors and needs to guess or extract them from what is observed in the state space. One of the typical choices in financial literature is to use lagged past returns \citep{val_mom_asness} as factors to predict future returns. We resort to a simple heuristic to select the best possible lagged variables by fitting the eq. (\ref{Eq:returns}) for a set of candidate lags. Then, we select the best one by minimizing the average  squared residuals.


Figure \ref{Fig:stud} and \ref{Fig:stud_ppo} showcase the average out of sample performances of DQN and PPO compared to the fully and partially informed benchmarks, when the return dynamics, driven by one mean-reverting factor, follows a Student's T with different degrees of freedom.

From Figure \ref{Fig:stud} we note that in the presence of extreme events (T-student noise), the DQN agents can control the trading costs and obtain equal or superior cumulative net $PnL$ with respect to the two  benchmark agents, especially for lower degrees of freedom where the misspecification has a greater impact, and extreme events are more frequent. The $SR$ fairly outperforms the benchmark towards the end of the training process in both cases. The RL agents learn how to control the higher amount of risk introduced in the environment, while model-based strategies like the benchmark should have considered this in advance. Figure \ref{Fig:stud_ppo} shows the same misspecified case for the PPO agent, which can  consistently manage the transaction costs and obtain higher net PnL than the benchmark, still exhibiting greater variance than DQN. We note that PPO performs better with respect to the benchmark when the latter is provided with partial information and needs to discover the persistence of the signal on its own.

The second proposed misspecification introduces heteroskedasticity in the asset returns by considering a GARCH process with $p=1$ and $q=1$ for the asset variance. For simplicity, we assume the predictable component of the returns to be an autoregressive model with a lag of order 1. 

\begin{figure}[tb]
      \centering\includegraphics{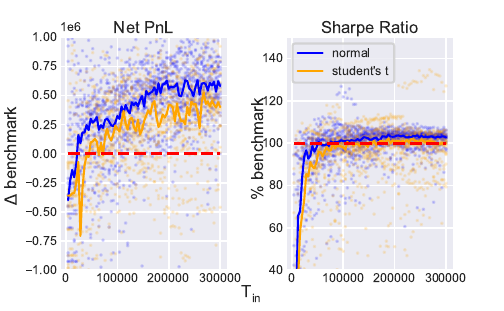}
      \caption{Results for AR-GARCH return dynamics when the noise is  distributed as a standard Normal or a Student's T distribution ($\nu=8$). The performance of DQN is compared with the dashed line benchmark. On the y-axis of the left plot, there is the cumulative net Pnl difference between DQN and the benchmark so the dashed line is set at 0. Then, the figure should be read with the same logic as Figure \ref{Fig:gauss} since the number of agents and the length of in-sample and out-of-sample tests are equal.}
      \label{Fig:garch}
\end{figure}

\begin{figure}[tb]
      \centering\includegraphics{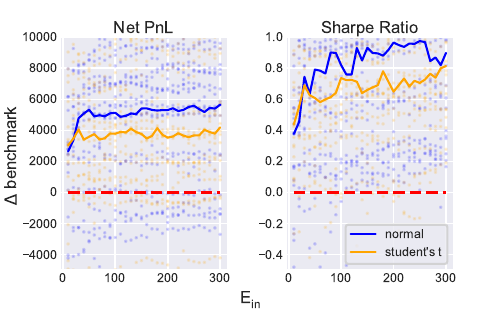}
      \caption{Results for AR-GARCH return dynamics when the noise is  distributed as a standard Normal or a Student's T distribution ($\nu=8$). The performance of PPO is compared with the dashed line benchmark. On the y-axis of both subfigures there is the difference between PPO and the benchmark, respectively for cumulative net PnL and the Sharpe ratio.  Then, the figure should be read with the same logic as Figure \ref{Fig:gauss_ppo} since the number of agents and the length of in-sample and out-of-sample tests are equal.}
      \label{Fig:garch_ppo}
\end{figure}

Figure \ref{Fig:garch} and Figure \ref{Fig:garch_ppo} present the evolution of DQN and PPO out-of-sample results as a function of the training set size,  when the return dynamics follows an AR-GARCH. Both figures present two noise specifications, such as a standard Normal or a Student's T distribution with $\nu=8$ degrees of freedom. Both figures should be read with the same logic as Figure \ref{Fig:gauss} since the number of agents and the length of in-sample and out-of-sample tests are equal.

Figure \ref{Fig:garch} shows that DQN obtains, on average, a higher cumulative net $PnL$ with respect to the benchmark. We compare the difference, instead of the ratio, between the two cumulative net $PnL$s because, in some cases, the net $PnL$ obtained by the benchmark agent is negative. Differently from the previous set of experiments, the increment of performance in the presence of heteroskedasticity regards the control of the amount of transaction cost. Looking at the $SR$, DQN mostly tracks the benchmark and outperforms it only in the case of Gaussian noises. The increased amount of extreme events in the Student's T case caused a worsening in the DQN performance relative to the benchmark. It has to be noted that we use the same set of hyperparameters for all these experiments. This is a signal that the performance in the fat-tailed case could be improved by tuning a more effective configuration. Figure \ref{Fig:garch_ppo} further confirms that PPO deals more effectively with model misspecifications with respect to DQN. When trained over GARCH dynamics, PPO achieves better than the benchmark performance in controlling associated risks and costs.

\begin{figure}[!htb]
      \centering\includegraphics{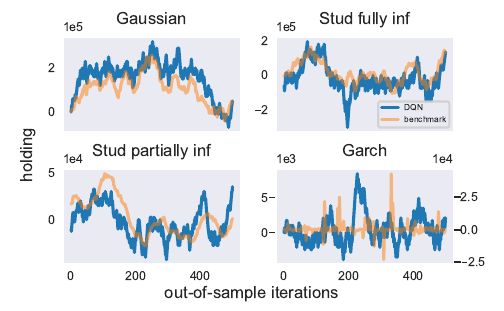}
      \caption{Portfolio Holdings for a snapshot of length 500 for some of the out-of-sample tests performed. The DQN agents selected are the best performing for each group in terms of cumulative net $PnL$: the Gaussian case with one factor, the Student's T case with 6 degrees of freedom (fully and partially informed) and the GARCH(1,1) with normal noise.}
      \label{Fig:holdings_dqn}
\end{figure}

\begin{figure}[!htb]
      \centering\includegraphics{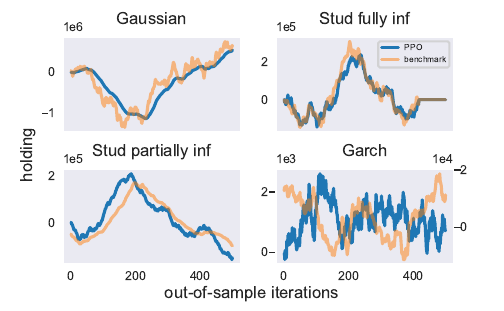}
      \caption{Portfolio Holdings for a snapshot of length 500 for some of the out-of-sample tests performed. The PPO agents selected are the best performing for each group in terms of cumulative net $PnL$: the Gaussian case with one factor, the Student's T case with 6 degrees of freedom (fully and partially informed), and the GARCH(1,1) with normal noise.}
      \label{Fig:holdings_ppo}
\end{figure}

Figure \ref{Fig:holdings_dqn} shows the realized out-of-sample holdings for some DQN agents. When the underlying dynamics can be predicted by  mean-reverting factors, as for the Gaussian and Student's T cases, the inversion of the factor sign causes the inversion of the sign of the portfolio itself. The oscillation between short and long positions confirms that the DRL algorithm has learned to follow the signal in the data. In particular, when compared with a partially informed benchmark agent, as in the bottom left plot of Figure \ref{Fig:holdings_dqn}, the DQN algorithm obtains a higher cumulative net $PnL$ by anticipating the reverting movement of the returns. Figure \ref{Fig:holdings_ppo} presents out-of-sample holdings for PPO in the same cases already shown for DQN. When the returns are driven by a Gaussian or a Student's T dynamics, PPO tracks the portfolio benchmark well and, in the partially informed case, seems to anticipate the mean-reversion of the signal as DQN does. In the case of GARCH dynamics, both the algorithmic approaches show a portfolio holding which greatly differs from the benchmark. We have two different $y$-axes for the bottom right panel in both figures. To visualize portfolio holdings that are on different scales, we let the left $y$-axis be associated with the algorithm and the right $y$-axis be associated with the benchmark. The latter does not adapt well to heteroskedastic peaks in the time series of simulated returns, and it happens that the benchmark portfolio is more sensitive to extreme events. On the other hand, RL can limit trading even in the presence of heteroskedasticity and obtain a lower portfolio size that produces fewer transaction costs when it needs to be rebalanced.

\begin{figure}[!htb]
      \centering\includegraphics{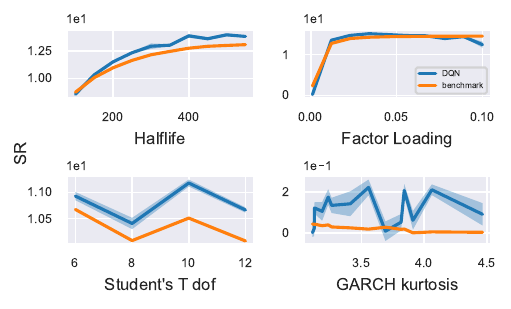}
      \caption{DQN performances measured by percentage or differential SR with respect to the benchmark when some relevant parameter of the simulated dynamics varies. The top row shows the one Gaussian mean-reverting factor dynamics when the level of the half-life of mean-reversion and factor loadings increases. The bottom row shows the one mean-reverting Student's T factor dynamics and the GARCH dynamics when respectively the degrees of freedom (dof) and the kurtosis of the returns distribution increase.}
      \label{Fig:robustness_dqn}
\end{figure}

\begin{figure}[!htb]
      \centering\includegraphics{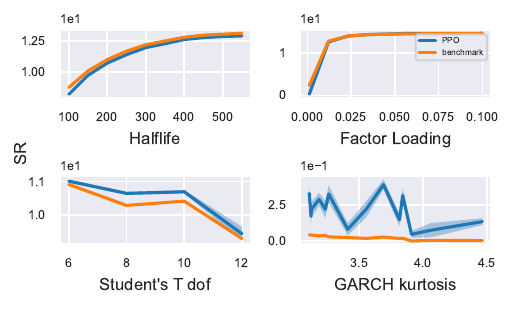}
      \caption{PPO performances measured by percentage or differential SR with respect to the benchmark when some relevant parameter of the simulated dynamics varies. The top row shows the one Gaussian mean-reverting factor dynamics when the level of the half-life of mean-reversion and factor loadings increases. The bottom row shows the one mean-reverting Student's T factor dynamics and the GARCH dynamics when respectively the degrees of freedom (dof) and the kurtosis of the returns distribution increase.}
      \label{Fig:robustness_ppo}
\end{figure}

It is important to check the robustness of the RL performance with respect to the choice of the dynamics parameters or, conversely its sensitivity w.r.t. some of them. Figures \ref{Fig:robustness_dqn} and \ref{Fig:robustness_ppo} show the level of SR for DQN and PPO agents as a function of dynamics half-life, factor loading, and fat-tailed return distribution parameters. The result is an average of over 10 different agents for each parameter configuration. This also allows to the creation of confidence intervals  around the average performance. In the top row of both figures, the effect of variation in the half-life of mean-reversion and in the factor loading $b$ of a one-factor Gaussian dynamics is similar for DQN and PPO. They both tend to obtain the same SR of the benchmark strategy, which is, in both cases, an increasing function. This is due by one side from the fact that a higher half-life produces a more persistent return sign that  agents can more easily exploit to make a profit. On the other side,  when the factor loading is small, then the  return dynamics contain no meaningful signal and the eq. (\ref{Eq:returns}) is driven purely by noise.  The left panel in the bottom row of \ref{Fig:robustness_dqn} and \ref{Fig:robustness_ppo} proposes the same sensitivity analysis when we simulate one factor Student's T dynamics with increasing degrees of freedom $\nu$. All the strategies get worse as the percentage of extreme events increases, but both PPO and DQN deal with riskier events in an effective way and consistently achieve a greater SR than the benchmark strategy. The right panel in the bottom row of both figures presents the variation of the performance when the kurtosis of the GARCH(1,1) process distribution increases. The fourth standardized moment for the stochastic volatility process in eq (\ref{Eq:garch}) is computed as
\begin{equation}
\frac{E\left(\sigma_{t}^{4}\right)}{\left[E\left(\sigma_{t}^{2}\right)\right]^{2}}=\frac{3\left[1-\left(\alpha_{1}+\beta_{1}\right)^{2}\right]}{1-\left(\alpha_{1}+\beta_{1}\right)^{2}-2 \alpha_{1}^{2}}
\end{equation}
knowing that when $1-2 \alpha_{1}^{2}-\left(\alpha_{1}+\beta_{1}\right)^{2}>0$, the tails in the distribution of the GARCH(1,1) are heavier than a Gaussian. A GARCH(1,1) model with heavy tails represents a consistent misspecification of the original conditions and only PPO consistently outperforms the benchmark while the tails of the simulated return distribution become thicker.


\FloatBarrier
\section{Conclusions}\label{conclusion}
In this work, we have used different RL algorithms to solve a trading problem in a financial environment where trading is costly. When the optimization problem is known to have an exact solution, DQN and PPO are able to track this benchmark, but they can also adapt to variations of the original environment setting and find a way to control portfolio risks and costs in a data-driven manner. While value-based DRL results are accurate in following the trading signals and controlling the market frictions, policy-based DRL appears to be more robust to extreme events and heteroskedastic volatility. 

Although DQN is able to learn the direction of the trades, the discretization of the action space still represents a major concern because the traded size is a multiple of a fixed quantity chosen in advance. Instead of moving towards actor-critic frameworks, which usually share instability issues with DQN, PPO helps in solving this issue by expressing the policy in a continuous action space.

 RL algorithms are demanding in terms of training data that can be quite scarce, especially at low frequencies. We believe that the use of a financial model with a  known optimal solution can offer a workaround to this problem. Classical strategies can facilitate the training of DRL agents by providing information about  good (even if  suboptimal) strategies. This can help  in terms of a rationalization of the state-action space, which results in the lightening of the training process itself.
 
 Moreover classical descriptive models  allow us to pre-train DRL agents on synthetic data and then fine-tune them on real-time series. Pre-training is possible thanks to the ability of model free DRL to find optimal strategies even in rich and realistic generative models
of financial markets, where time and data scarcity is not an issue.  Fine-tuning could be done by using a residual approach (see Appendix \ref{sec:RL}). This is an interesting scenario in which model-based and model-free approaches can collaborate. It could be interesting for future research to investigate also the possible use of Residual Reinforcement Learning as an operational measure of the goodness-of-fit of a model to data. For example, instead of the classical likelihood or moment-based approaches, one could consider a model to be representative of the real market if the model free RL agent (trained on data) acts similarly to the one trained on the model, i.e., if the residual actions are small.  

Finally, it would be also interesting to investigate the generalization properties of RL when trained on heterogeneous time series.  In real applications, one
typically would want to train the agent using all the data from all the available stocks, hoping that it will be able to trade effectively on every single stock, not just because it has learned a single strategy that is good in average but because  it can associate to each different stock the associated corresponding strategy. This is similar to a long training in a period with different price regimes hoping that the agent will be able to act in the future, adapting its strategy to the presence of regime switching and a certain level of non-stationarities.

\section*{Acknowledgments}
DT acknowledges GNFM-Indam for financial support. This work was partially supported by project SERICS (PE00000014) under the MUR National Recovery and Resilience Plan
funded by the European Union - NextGenerationEU. 

\bibliography{ms}

\begin{appendices}

\section{Algorithms and Hyperparameters}
In this section,  we provide some details regarding the implementation of Q-learning and DQN algorithms used in the numerical  experiments. Then we outline the choice of the parameters for simulating the financial data and of the relevant hyperparameters to set up the training of the algorithms. 
\subsection{Q-learning}
Q-learning requires the discretization of the state space $\mathcal{S}$ and the action space $\mathcal{A}$, which affects the dimensionality of the Q-table that produces estimates $Q(S_{t},A_{t})$ of the optimal action-value function. For every possible state and action variable, we need to choose a proper discrete range that we believe is adequately large to capture the relevant information and solve the problem. 

Since real traders usually operate by trading quantities of assets that are multiples of a fixed size called lot, the dimensions of the Q-table are bounded by setting the traded quantity $\Delta h_{t}$ to be at most K round lots and the portfolio holding $h_{t}$ to a maximum of M round lots. The discrete set of returns is represented by an upper and lower bounded set of values that are linearly spaced by the size of a basis point denoted as $bp$. The bounded sets and their dimensionality are respectively:
\begin{align}
    & \mathcal{A} =  \left\{ -K,-K+1, \ldots, K \right\}, \quad \Bigl\vert \mathcal{A} \Bigr\vert = 2K + 1 \\
    & \mathcal{H} = \left\{ -M,-M+1, \ldots, M \right\}, \quad \Bigl\vert \mathcal{H} \Bigr\vert = 2M + 1 \\
    & \mathcal{R} = bp \cdot \left\{ -T,-T+1, \ldots, T \right\}, \quad \Bigl\vert \mathcal{R} \Bigr\vert = 2T + 1
\end{align}
In our financial environment, the basis point and the lot are respectively the size of minimum return movement and the minimum tradable quantity of the asset at each discrete time. The sizes of the three sets are defined respectively by the hyperparameters $K$, $M$ and $T$, which are a crucial choice to define the magnitude of the synthetic financial problem. 

Denoting the size of the table as $d = \bigl\vert \mathcal{R} \bigr\vert \times \bigl\vert \mathcal{H} \bigr\vert  \times \bigl\vert \mathcal{A} \bigr\vert$, our simulated experiments show that Q-learning is not even able to reach a positive profit when $d$ approaches the length of the simulated series $T_{in}$. The more the dimensionality of the table increases, the worse are the cumulative net PnLs and rewards obtained, when the $T_{in}$ is fixed. 

If $T_{in}$ is not sufficiently long to allow the agent to visit the entire state space and  update the Q-table in each corresponding entry, the algorithm approximates the action-value function with a sparse Q-table. Thus, it is not able to represent the effect of slight changes in the state space variables. Such a bottleneck becomes even worse if we increase the number of actions that the agent can perform. 

In principle, we could let Q-learning experience longer simulated series to partially avoid the exploration issue, but this would not be of any practical use for two reasons: (i) it requires an increasingly long training runtime to match the benchmark performance and still this result would be obtained under a discretized state-space of a more complex financial environment; (ii) training for a high number of iteration the experiment would not even resemble a real financial application since there is no way to retrieve such a massive amount of financial data, especially at a daily frequency. 

To ensure the proper exploration of the state space, the agent acts according to an $\epsilon$-greedy policy, such that at each time a greedy action $a=\argmax_{a} Q(S_{t}, a )$ is selected with probability $1-\epsilon$, while occasionally with probability $\epsilon$ a random action is sampled from the set $\mathcal{A}$. As a common approach in the RL literature, the value of $\epsilon$ decays linearly during training until it reaches a small value that is kept fixed until the end.

\subsection{DQN}
DQN requires a discretization of the action space $\mathcal{A}$, which is approached as for the tabular case. This discretization could represent an issue when one wants to represent the choice of the agent at a more granular level. The more one increases the size of $\mathcal{A}$, the more the computational cost of the algorithm increases, and its efficiency in solving the financial problem decreases. However, we believe the discrete control can still be adequate for a set of financial problems since usually market orders are executed in multiples of a fixed quantity. 

Since the agent learns offline by choosing past batches of experience from a buffer with a fixed size, we set this dimension as a percentage of the total updates in sample $T_{in}$. We have found that letting the buffer size increase can improve the performance, therefore we do not discard any sequence. The exploration-exploitation trade-off is balanced as in Q-learning, using an $\epsilon$-greedy policy where the $\epsilon$ decreases linearly to a low value towards the end of the training. Despite the original DQN implementation \citep{mnih2015human} suggests updating the target network parameters at every fixed discrete step, we choose to continuously update the target parameter so that they slowly track the learned networks as follows:
\begin{equation}
   \theta^-_t \longleftarrow \tau \theta^-_t +(1-\tau) \theta_t
\end{equation}
where $\tau$ is the chosen step size for the update, $\theta^-_t$ are the target network parameters and $\theta$ are the parameters for the current action-value function estimator. 

A problem of the overestimation of the action-value function is known to arise in the classical DQN algorithm \citep{Hasselt10,hasselt2015doubledqn}. Thus we adopt the double DQN (DDQN) variant suggested in \cite{hasselt2015doubledqn}. Recalling that the target of a DQN update is computed as
\begin{equation}\label{tgt_dqn}
    T^{\text{DQN}}_t=R_{t+1} + \rho \max_{a}Q(S_{t+1},a;\theta),
\end{equation} 
where we can write $$\max _{a^{\prime}} Q\left(s^{\prime}, a^{\prime} ; \theta\right) = Q\left(s^{\prime}, \argmax_{a^{\prime}}Q\left(s^{\prime}, a^{\prime} ; \theta\right) ; \theta\right),$$
it happens that the same noise affects both the maximization over the action space and the value function estimates. Removing the correlation between the sources of noise coming into these two operations is beneficial to avoid an overestimation of the value function. 

Double Q-learning decouples the selection of the action from the evaluation as 
\begin{equation}\label{DDQN}
    T^{\text{DDQN}}_t = R_{t+1} + \rho Q(S_{t+1},\argmax_{a^{\prime}} Q\left(S_{t+1}, a^{\prime} ; \theta_{1}\right) ;\theta_2),
\end{equation}
i.e., DDQN uses two neural networks: one computes the target, and the other computes the current action-value function. The computation of the target is split between the current neural network that greedily selects the action and the target neural network that evaluates such action. Therefore, the selection of the action in eq. (\ref{DDQN}) is due to the current weights $\theta_1=\theta_{t}$, while the target network is used to evaluate the value function for that action $\theta_2=\theta^-_t$. 

We have found that DDQN outperforms DQN in all the tests we carried out, meaning that also, in this specific financial application, it is a profitable procedure.

Regarding the shape of the loss function, a common choice for the DQN family of algorithms is the Huber loss rather than the mean squared error (MSE), which is typical for regression tasks. Huber loss is less sensitive to the presence of  outliers, and it is expressed as
\begin{equation}
\mathcal{L}_{\delta}(y, \hat{y})=\left\{\begin{array}{ll}
\frac{1}{N} \sum_{i=1}^{N} \left(y_{i}-\hat{y_{i}}\right)^{2} & \text { for }\mid y_{i}-\hat{y}_{i}\mid  \leq \delta \\
\delta\frac{1}{N} \sum_{i=1}^{N}\mid y_{i}-\hat{y}_{i}\mid -\frac{1}{2} \delta^{2} & \text { otherwise }
\end{array}\right.  
\end{equation}
The Huber loss is quadratic for small values of the squared difference and linear for larger values. This kind of loss function adds a lower penalty to large errors and it is better than MSE for this kind of problem because learning by estimates as in DQN could produce unexpectedly high errors. Even if in the main we showed the case of an MSE loss, in practice our implementation utilizes a Huber loss. This choice does not change the update rule of the presented algorithms, but the computation of the gradient will differ in the presence of large quadratic errors.

The neural networks used to approximate the action-value function are 2-layer fully connected networks with ELU \citep{clevert2015fast} activation  and uniform weight initialization as in \cite{he2015delving}. We have tried different types of rectified nonlinear activation, but ELU outperforms a more usual choice as ReLU. The sizes of the hidden layers are 256 and 128 for the first and the second respectively, but also smaller hidden layers have proven to be effective \citep{xu2015empirical}. 

The gradient descent optimizer is Adam \citep{kingma2014adam} which performs a batch update of size 256. The original implementation proposes default values for $\beta_{1}$, $\beta_{2},$ and $\epsilon^{\text{adam}}$, which are respectively the exponential decay rates for the first and the second-moment estimates of the true gradient and a small constant for numerical stability. Those parameters required some tuning for improving performances so we set them as $\beta_{1}=0.5$, $\beta_{2}=0.75$ and $\epsilon^{\text{adam}}=0.1$ for all experiments. The learning rate $\alpha$ usually starts around 0.005 and then decays exponentially  towards the end of the training.

Since in an RL setting the data are not all available at the beginning of the training, we can not normalize our input variables as usual in the preprocessing step of a supervised learning context. Hence, we add a Batch Normalization layer \citep{ioffe2015batch} before the first hidden layer to normalize the inputs batch by batch and obtain the same effect. 

\subsection{PPO}
PPO allows the expression of continuous policies through an algorithm which is easier to implement than a trust-region method \citep{trpo2015} and easier to tune with respect to the continuous counterpart of DQN \citep{LillicrapHPHETS15}. In principle, continuous policies are more expressive than discrete policies but are also harder to learn. Our implementation of PPO follows \cite{andrychowicz2020matters} which performs a large empirical study of the effect of implementation and parameter choices on the PPO performances. Even if our financial problem differs from their testbed, we also follow the direction of their results in order to tune our hyperparameters since we have limited computational resources to do this search from scratch. Another relevant source for effective implementation is  \cite{engstrom2020implementation}.

As described in the main, we implement PPO in an actor-critic setting without shared architectures. Differently from the standard implementation, the actor outputs a single scalar value, which is the mean of a Gaussian distribution, while the standard deviation is then learned as a global parameter in the optimization process and updated using the same gradient optimizer. Learning a global standard deviation for all the state representations has proven to be as much as effective in learning a state-dependent parameter, with the benefit of being slightly less computationally expensive \citep{andrychowicz2020matters}. Policy gradient methods like PPO allow inserting some prior knowledge on the form of the policy with respect to value-based methods. The real-valued output of the actor usually passes through a hyperbolic tangent function in order to bound the action in the interval $[-1,1]$. Then is rescaled to directly express a range of possible trading actions whose extreme values are selected according to a heuristic described in the next subsection. Exploration during training is guaranteed by the learned standard deviation parameter and by the entropy bonus in the objective function. When doing out-of-sample tests, the PPO policy is tested as if it were deterministic by just picking the mean of the Gaussian instead of sampling from it. We made this choice because we do not want the test results to be affected by stochasticity.

The on-policy feature of PPO makes the training process episodic, so that experience is collected by interacting with the environment and then discarded immediately once the policy has been updated. The on-policy learning appears, in principle a more obvious setup for learning even if it comes with some caveats because it makes the training less sample efficient and more computationally expensive since a new sequence of experiences needs to be collected after each update step. In this process, the advantage function is computed before performing the optimization steps, when the discounted sum of returns over the episode can be computed. In order to increase the training efficiency, after one sweep through the collected samples, we compute the advantage estimator again and perform another sweep through the same experience. This trick reduces the computational expense of recollecting experiences and increases the sample efficiency of the training process. Usually, we do at most 3 sweeps (epochs) over a set of collected experiences before moving on and collecting a new set.

The optimizer and the normalization of the inputs through a Batch Normalization layer are the same used for DQN, with the only exception that in the PPO case, we do not tune the hyperparameters of the Adam optimizer.

Maximizing the objective function that returns the gradient in eq. (\ref{Eq:pgadvantage}) is known to be unstable since updates are not bounded and can move the policy too far from the local optimum. Similarly to TRPO \citep{trpo2015}, PPO optimizes an alternative objective to mitigate the instability
\begin{equation}\label{Eq:CLIP}
J^{\text{CLIP}}(\theta,\psi) = \mathbb{E}_{\pi_{\theta}} \left[ \min \left( r(\theta) \hat{\mathbb{A}}\left(s,a;\psi\right),  \operatorname{clip}\left(r(\theta), 1-\epsilon, 1+\epsilon\right) \hat{\mathbb{A}}\left(s,a;\psi\right) \right) \right]
\end{equation}
where $r(\theta) = \frac{\pi\left(A_{t} \mid S_{t}; \theta\right)}{\pi\left(A_{t} \mid S_{t}; \theta_{\text {old}}\right)}$ is a ratio indicating the relative probability of action under the current policy with respect to the old one. Instead of introducing a hard constraint as in TRPO, the ratio is bounded according to a tolerance level $\epsilon$ to limit the magnitude of the updates. The combined objective function in eq. (\ref{Eq:ppoobj}) can be easily optimized by PyTorch's automatic differentiation engine, which quickly computes the gradients with respect to the two sets of parameters $\theta$ and $\psi$. The implemented advantage estimator depends on the parametrized value function $V_{\psi}$ and is a truncated version of the one introduced by \cite{mnih2016asynchronous} for a rollout trajectory (episode) of length $T$:
\begin{equation}
\hat{\mathbb{A}}_{t}=\delta_{t}+(\gamma \tau) \delta_{t+1}+\cdots+\cdots+(\gamma \tau)^{T-t+1} \delta_{T-1} 
\end{equation}
where $\delta_{t}=r_{t}+\gamma V_{\psi}\left(s_{t+1}\right)-V_{\psi}\left(s_{t}\right)$, $\gamma$ is a discount rate with the same role of $\rho$ in DQN and $\tau$ is the exponential weight discount which controls the bias-variance trade-off in the advantage estimation. The generalized advantage estimator (GAE) uses a discounted sum of temporal difference residuals similar to the one-step target value of DQN in eq. (\ref{Eq:DQNtgt}).


\subsection{Environment choices}
In all the simulated experiments we set $\bigl\vert \mathcal{A} \bigr\vert = 5$ for both Q-learning and DQN so that every agent can perform two buy actions, two sell actions, and a zero action. In order to make the results comparable with those of the benchmark solution, we adopt a systematic way to choose the size of the action space $\mathcal{A}$ from which we obtain the possible actions. Basically, we let the dynamic programming solution run for some iterations before starting the training of the RL algorithms, and we look at the distribution of the continuous action performed. Then we select the lower and upper boundary of $\mathcal{A}$ as respectively the quantiles located at the 0.1\% and the 99,9\% of that distribution. In doing so, we avoid extreme actions of the benchmark agent and allow the RL agents to operate in the most possible similar setting to be able to compare the performances. We could adopt the same approach for the discretization of the variable in $\mathcal{S}$ as required by Q-learning, but this would produce very large Q-tables ending up being very sparse after the training process. Therefore, for Q-learning, we choose $T=0.05$ and $M=100000$.

For what concerns the simulated environment, the cost multiplier $\lambda$ is chosen to equal 0.001 for all experiments. Then we assume a zero discount rate and all the starting positions for the holding are zero. For any experiment involving mean-reverting factors, we compute the speed of mean reversion $\phi$ as $\phi = \frac{\log(2)}{\log(h)}$ where $h$ is known as the half-life of mean-reversion and represents the time it is expected to take for half of the trading signal to disappear. This allows us to simulate the predicting factors and aggregate their effect to compute the asset returns. We tried many different sets up for the half-lives, factor loadings, and volatility of simulated assets and the findings are quite robust. The half-lives of the mean-reverting factors are 350 for the case of a single factor and (170,350) for the case of two factors. Factor loadings are chosen of the order of magnitude of $10^{-3}$, specifically as 0.00535 and 0.005775 in the proposed cases. The volatilities of the factor are respectively 0.2 and 0.1, while the volatility of the unpredictable part of the asset return is always set to 0.01. Suitable ranges for these hyperparameters is $[0.5,0.05]$ for the former and $[0.05,0.005]$ for the latter.  In general, DQN is also able to retrieve the underlying dynamics in the case of two concurrent factors with different speeds, as long as those factors do not include one which is really fast (e.g. half-life of mean reversion lower than 10 days) and also highly noisy with volatility above 0.2. This is acceptable because the signal-to-noise ratio would be very low and it may require more sophisticated layers for feature extraction, other than a feedforward network structure. The parameters for the AR-GARCH simulation are $\omega=0.01$, $\alpha_{1}=0.05$, and $\beta_{1}=0.94$ which are common GARCH parameters to simulate a stable financial market. The autoregressive parameter is set to $ \phi_{L_{r}}=0.9$, and the degrees of freedom in the Student's T case are $\nu=10$.



\subsection{Algorithm stability}
Despite the great empirical success of deep reinforcement learning, its theoretical foundation is less well understood. Very often, the most widely used algorithms that are effective in practical applications are exactly those lacking theoretical guarantees.

To our knowledge, the dominating framework for most of the theoretical results to model the interaction between agent and environment is the Markov decision process (MDP), which in our context translates into a Markovian process for price dynamics.

In this framework, the optimal value function is characterized as a fixed point of the Bellman operator. A fundamental result
for MDP is that the Bellman operator is a contraction in the
value-function space, so the optimal value function is the
unique fixed point. Furthermore, starting from any initial
value function, iterative applications of the Bellman operator ensure convergence to the fixed point (see \cite{puterman2014markov} for details.) Many of the most effective RL algorithms have their root in such a fixed-point view. The most prominent family of
algorithms is perhaps the temporal-difference algorithms, including  Q-learning (\cite{watkins1989learning}) and DQN (\cite{mnih2015human}).

\subsubsection*{Q-learning}
When the Bellman operator can be computed exactly (even
on average), such as when the MDP has finite state/actions,
convergence is guaranteed thanks to the contraction property. This is the case for tabular Q-learning as explored in \cite{watkins1992q} first and \cite{melo2001convergence} then: given a finite MDP the Q-learning algorithm is convergent to the optimal Q-function provided that the learning rate $0\leq\alpha_t<1$ satisfies 
$$ \sum_t \alpha_t =\infty, \ \ \ \ \sum_t \alpha^2_t<\infty$$
and that the exploration policy $\pi$ satisfies $\pi$ , for every state-action pair $(s,a)$. This means that 1) the learning rate, which is the parameter used to scale the value function update, should be large enough to ensure adequate exploration and small enough not to prejudice convergence, and 2) the agent is trained off-policy such that it must visit each possible state-action pair (it will do it an infinite number of times).  

Q-learning and tabular algorithms, in general, are known to have severe drawbacks in practice, pointing towards the curse of dimensionality problem. When the number of possible state-action pairs is too large, or the MDP is not finite/discrete and requires a proper discretization, the algorithm becomes computationally expensive in terms of memory and convergence time. This constitutes the primary motivation for introducing the concept of function approximation for reinforcement learning.

\subsubsection*{DQN}

When function approximations, such as neural networks, are used, a theoretical analysis is highly challenging, the success is actually empirical, and the performance relies heavily
on hyperparameter tuning. This happens because, in temporal difference algorithms with function approximations, the agent uses its own prediction to construct the learning objective (i.e., bootstrapping). Since predictions over different data interfere with each other, the learning objective is unstable
during training and potentially leads to divergence (see \cite{baird1995residual}). Convergence is guaranteed only in a few special cases:  some rather restrictive function classes, such as
those with a non-expansion property (see \cite{gordon1995stable,ormoneit2002kernel,antos2008learning}); in certain specific cases with linear function approximators (see \cite{sutton2009fast,maei2010toward} ); for just policy evaluation (see for example \cite{boyan2002technical}). Other convergent gradient-based methods have been proposed (\cite{feng2019kernel,ghiassian2020gradient}), but they cannot easily be used with deep non-linear neural networks as they involve computationally heavy operations, and they often show worse empirical
performance.

In the case of DQN, the non-convergence problem has been empirically investigated by \cite{van2018deep}. A theoretical analysis
of DQN is highly challenging because, in addition to the aforementioned  Q-network approximation leading to instabilities,
it uses another neural network named the target network to obtain an unbiased estimator of the mean-squared Bellman error used in training the Q-network. The two networks are synchronized and still coupled. However, even if we fix the target network and focus on updating the Q-network, the subproblem of training a neural network still remains less well-understood in theory. To our knowledge, the closest steps towards a stability control of DQN resides in \cite{fan2020theoretical} and \cite{wang2021convergent}. In the first paper, DQN is proved to be convergent under some simplifications and unpractical assumptions: a ReLu network activation function is used to exploit their approximation properties; it 
is assumed that every training set of reward-state transitions is drawn, i.i.d. and that a global minimum of the Q-function on the training set is provided. The second paper proposes the DQN (C-DQN) modification based on different loss functions and tested on standard benchmark problems. In this case, stability is guaranteed (only in the sense that the loss monotonically decreases).

\subsubsection*{PPO}
Proximal Policy Optimization (PPO) algorithm, which is not a temporal difference method but a policy gradient, actor-critic method, also lacks theoretical guarantees. In fact, actor-critic algorithms have been only proven to converge for simple settings like neural networks that are linear (\cite{konda1999actor,yang2019provably,liu2019neural}). The main difficulties of dealing with PPO are that 1) it uses episodes as samples instead of transitions, 2) it uses policies that become greedy (thus poor exploration), and  3) its objectives use previous policies estimation (trust region methods). 
Recently \cite{holzleitner2021convergence} proved PPO convergence on finite MDP using the two time-scales stochastic approximation theory (\cite{borkar2008stochatic}).
However, while a framework to ensure convergence is developed, it does not imply convergence to an optimal policy.
Such proofs are, in general, difficult for methods that use deep neural networks since locally stable
attractors may not correspond to optimal policies (\cite{jin2020local,lin2020gradient}).\\

\subsection{Hybrid framework}\label{sec:RL}

Model-free RL algorithms and model-based approaches may work synergistically in bidirectional feedback. An RL agent can find close-to-optimal strategies in rich and non-gaussian financial market models, overcoming the difficulties of a model-based approach ( analytical approximations and estimation errors) and indirectly measuring the relevant features of the price dynamics. At the same time, an RL agent can leverage prior knowledge of a model-based classical agent that can improve its efficiency and reduce the amount of experience needed for convergence.  

To deepen this idea, we also test what is generally called a \textit{Residual Learning} approach. 
Specifically, we express the quantity to trade at each time step as a residual part 
\begin{equation}\label{Eq:BKresidualtrade}
    \Delta h_{t} = \Delta h_{t}^{g}(1 - a_t) 
\end{equation}
where $h_{t}^{g}$ is a gaussian reference portfolio, which one can choose to be the Markowitz or the Garleanu-Pedersen  portfolios,
and $a_t$ is a real-numbered action between 0 and 1. The action $a_t$  expresses the percentage of the reference portfolio rebalance, which is better not to buy for the current time $t$ to contain the costs, to take into account estimation errors for $h_{t}^{g}$ and considering possible deviations from the gaussian price. The reinforcement learning agent aims to learn how to trade a residual version of the reference portfolio: a fine-tuning, thus, hopefully, an easier task.

In the Gaussian context, we know from \cite{GarPed13} that the trader needs always to slow down the trade with respect to what is suggested by the Markowitz solution. It is equivalent to saying that the agent learns to move toward the reference portfolio at a rate defined by the decay of the signal. If the decay is too fast, the agent will slow down the trade because the profitable opportunity is not supposed to be durable. On the contrary, if the decay is slow, the agent trades more closely to the reference since the profitable opportunity is long-lasting. The impact of this whole mechanism is lowering the price paid to rebalance the portfolio at every period with respect to an approach that does not account for the effect of the frictions, which we know from \cite{gueant2013permanent,patzelt2018universal,kyle2016market} to be highly nonlinear.

\begin{figure}[!htb]
     \centering
     \includegraphics[scale=0.8]{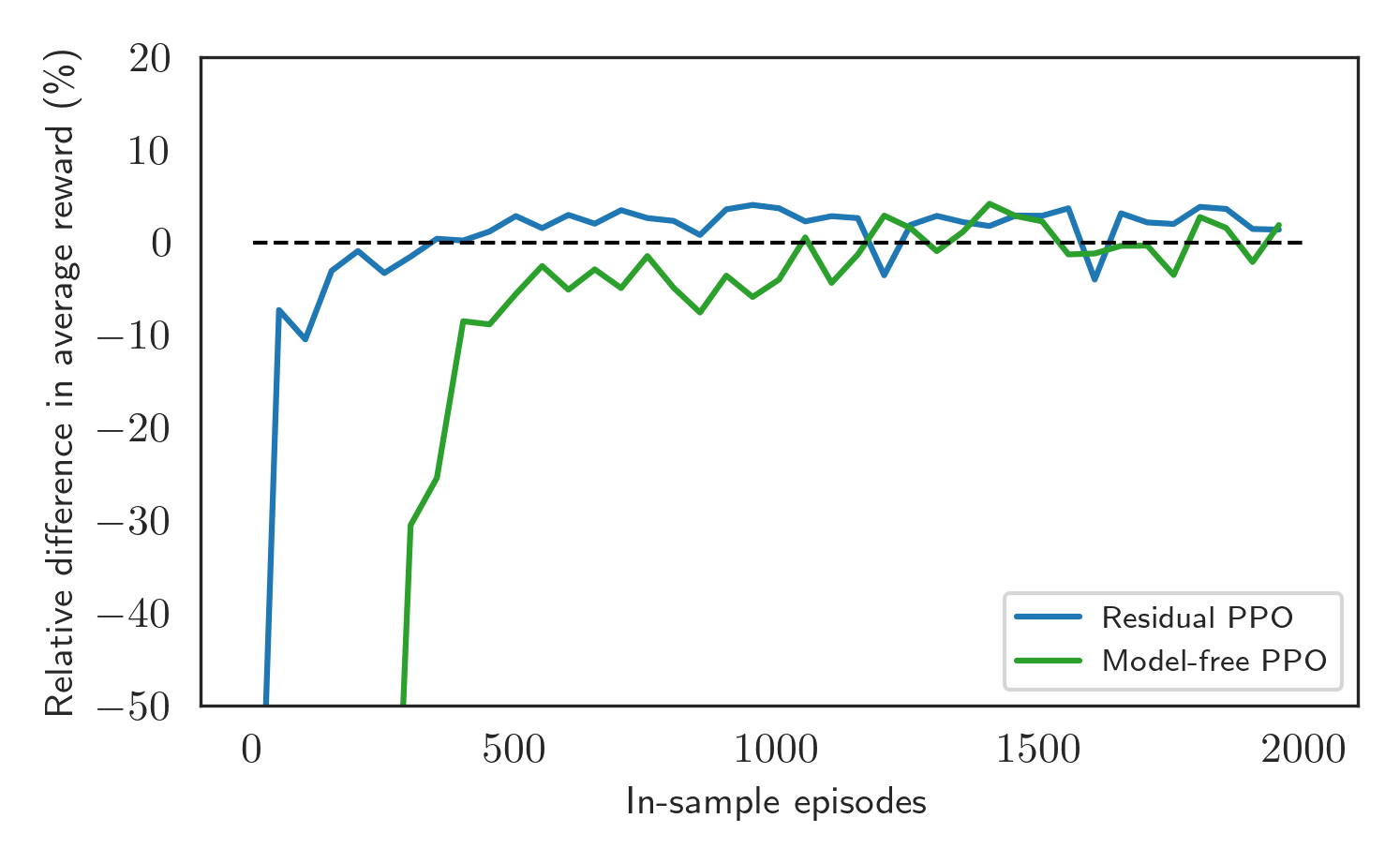}\\
     \caption{Convergence to the benchmark solution for the residual ($h^g_t$ is Markovitz) and the model-free reinforcement learning approaches trained using PPO with synthetic Student's T returns (8 degrees of freedom) driven by one-mean reverting factor. For each approach, we train 20 agents and calculate their average cumulative reward in-sample to choose the best-performing one. Then a rolling average of the relative difference in reward with respect to the benchmark is computed to compare them. }
     \label{Fig:convergenceres2}
\end{figure}

The residual approach's first advantage is shown in Fig. \ref{Fig:convergenceres2}. It is evident how the residual RL approach converges faster to a stable solution. On the other hand, the completely model-free approach requires more training episodes to reach in-sample performance stability.
\FloatBarrier
\FloatBarrier

\begin{figure}[H]
  \centering
  \includegraphics[scale=0.65]{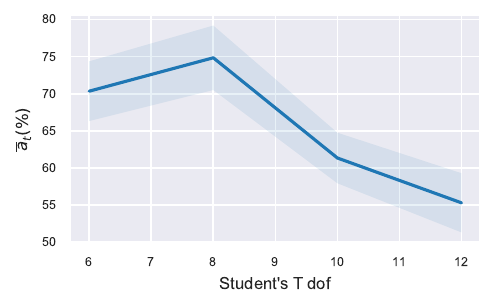}
  \includegraphics[scale=0.65]{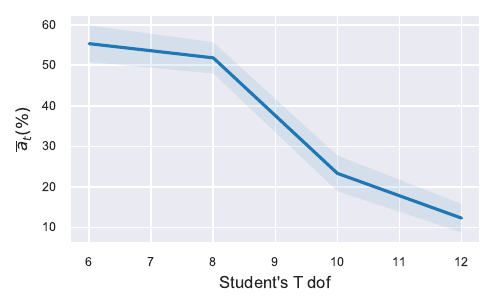}
  \caption{Residual learning average action $\bar{a_t}$ of a representative DQN (left) and PPO (right) agent, when the price dynamics has Student's T extreme events, as a function of the degrees of freedom. The reference strategy $h^g_t$ is Markovitz.}
  \label{Fig:distance_stud_1}
\end{figure}

\begin{figure}[H]
  \centering
  \includegraphics[scale=0.65]{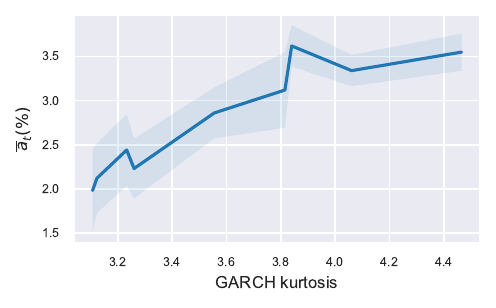}
  \includegraphics[scale=0.65]{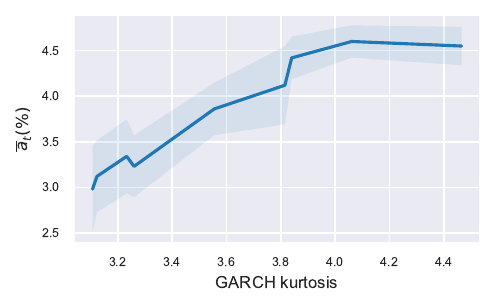}
  \caption{Residual learning average action $\bar{a_t}$ of a representative DQN (left) and PPO (right) agent, when the price dynamics follows a AR-GARCH, as a function of the effective model kurtosis.}
  \label{Fig:distance_garch_1}
\end{figure}

The second advantage is that the Residual approach gives us direct information about the number of non-gaussian behaviors present in the price dynamics without any model and parameter estimation. In fact, we can interpret the action $a_t$ as the distance from  the (gaussian) reference strategy and, therefore, indirectly, the distance of the price dynamics from the efficient market hypothesis. 
Fig. \ref{Fig:distance_stud_1}, for example, shows the average action $\bar{a_t}$ of agents trained using the residual learning approach on price dynamics with (T-student) extreme events. It is evident how, for both DQN (left) and PPO (right), the $\bar{a_t}$ is a decreasing function of the degrees of freedom, being a measure of the distance between the learned and the gaussian reference strategy which is not optimal in this market. 
Note that the specific value of $\bar{a_t}$ seems to depend on the RL algorithm since they exploit alternative strategies, but the qualitative behavior is the same. Similarly, Fig. \ref{Fig:distance_garch_1} shows the behavior of $\bar{a_t}$ when the underline price dynamics present volatility clustering. As expected in this case, $\bar{a_t}$ increases with the GARCH kurtosis (chosen as a metric of the amount of nongaussianity in the model).
In some sense, the average action of the Residual RL agent could be interpreted as an operational definition of non-Gaussianity in the financial market.

\end{appendices}

\end{document}